\documentclass{ws-p10x7}

\begin{document}

\title{THE MIXING OF THE $f_0(1370)$, $f_0(1500)$ AND
$f_0(1710)$ AND THE SEARCH FOR THE SCALAR GLUEBALL}

\author{A. Kirk}

\address{School of Physics and Astronomy,
Birmingham University, U.K.\\
E-mail: ak@hep.ph.bham.ac.uk}

\twocolumn[\maketitle\abstract{
For the first time a complete data set
of the two-body decays of
the $f_0(1370), f_0(1500)$ and $f_0(1710)$ into all pseudoscalar mesons
is available.
The implications of these data
for the flavour content for these three $f_0$ states is studied.
We find that they are in accord with the hypothesis
that the scalar glueball
of lattice QCD mixes with the $q \overline q$ nonet that also exists
in its immediate vicinity. We show that this
solution also is compatible with the relative production
strengths of the $f_0(1370), f_0(1500)$ and $f_0(1710)$
in ${\rm pp}$ central production, ${\rm p \bar{p}}$ annihilations
and $J/\psi$ radiative decays.
}]

\section{Introduction}%1
It is now generally accepted that
glueballs will mix strongly
with nearby $q \overline q$ states with the same
$J^{PC}$ and that
this will lead to three isoscalar states of the same
$J^{PC}$ in a similar mass region.
In general
these mixings will negate the naive
folklore that glueball decays would be
``flavour blind ".
\par
Lattice gauge theory calculations (in the quenched approximation)
predict that the lightest
glueball has $J^{PC}$~=~$0^{++}$ and that its
mass is in the $1.45-1.75$~GeV region. This
means that the three states in the glueball mass range are
the $f_0(1370)$, $f_0(1500)$ and the $f_0(1710)$.

\section{Data from the WA102 experiment}%2
\begin{figure*}
\epsfxsize24pc
\figurebox{}{}{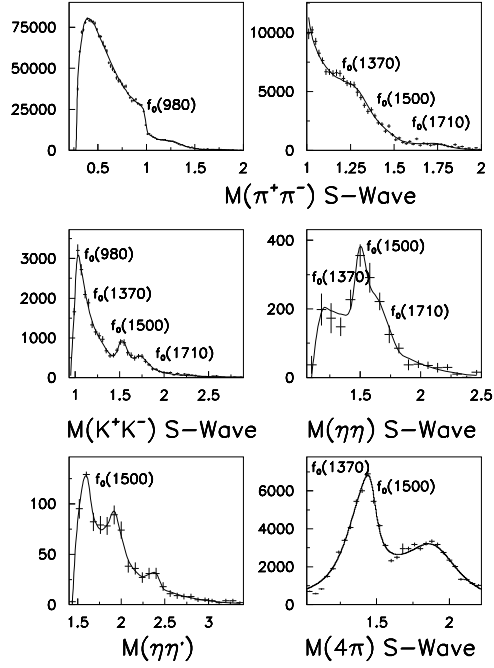}
\caption{The observation of the $f_0(1370)$, $f_0(1500)$ and
$f_0(1710)$ in the WA102 experiment.}
\label{fi:1}
\end{figure*}

Recently
the WA102 collaboration has published~\cite{etaetapap},
for the first time in a single experiment, a complete data set
for the decay branching
ratios of the
$f_0(1370)$, $f_0(1500)$ and
$f_0(1710)$ to all pseudoscalar meson pairs (see fig.~\ref{fi:1}).
\par
A coupled channel fit to this data yields
sheet II pole positions of
M($f_0(1370)$) = (1310 $\pm$ 19 $\pm$ 10)  $-i$  (136 $\pm$ 20 $\pm$ 15)  MeV
M($f_0(1500)$) = (1508 $\pm$ 8 $\pm$ 8) $-i$ ($54$ $\pm$ $7$ $\pm $ 6) MeV and
M($f_0(1710)$) = (1712 $\pm$ 10$\pm$ 11) $-i$ ($62$ $\pm$ $8$ $\pm $ 9)  MeV.
\par
The relative decay rates
$
\pi \pi : K \overline K : \eta\eta : \eta\eta^\prime : 4\pi
$
are for the
$f_0(1370)$:
$
1 : 0.46 \pm 0.19 : 0.16 \pm 0.07 : 0.0 : 34.0 ^{+22}_{-9}
$
for the
$f_0(1500)$:
$
1 \;:\; 0.33 \pm 0.07\;
:\; 0.18 \pm 0.03\; :\; 0.096 \pm 0.026 \;
:\; 1.36 \pm 0.15
$
 and for the
$f_0(1710)$:
$
1 : 5.0 \pm 0.7 : 2.4 \pm 0.6
: <\;0.18\;(90\;\%\;\; CL) : <\;5.4\;(90\;\%\;\; CL)
$

\par
These data will be used as input to a fit to
investigate
the glueball-quarkonia content of the $f_0(1370)$, $f_0(1500)$ and
$f_0(1710)$.

\section{The fit}%3

\par
In the $|G\rangle=|gg\rangle$, $|S\rangle=|s\bar{s}\rangle$,
$|N\rangle=|u\bar{u}+d\bar{d}\rangle/\sqrt{2}$ basis,
the three physical states can be read as
\begin{equation} \left( \begin{array}{ccc}
|f_0(1710)\rangle\\
|f_0(1500)\rangle\\
|f_0(1370)\rangle
\end{array}\right)
=\left(\begin{array}{ccc}
x_1 & y_1& z_1\\
x_2& y_2& z_2\\
x_3 & y_3 & z_3
\end{array}\right)
\left(\begin{array}{ccc}
|G\rangle\\
|S\rangle\\
|N\rangle
\end{array}\right),
\label{eq:e}
\end{equation}
where the parameters $x_i$, $y_i$ and $z_i$ are related to the
partial widths of the observed states~\cite{scalars}
as given in table~\ref{ta:1}.
\begin{table}[h]
\caption{The theoretical reduced partial widths.}
\label{ta:1}
\vspace{-0.1in}
\begin{center}
\begin{tabular}{|c|c|} \hline
 &  \\
$\gamma^2(f_i\rightarrow \eta\eta^\prime)$
&$2[2\alpha\beta(z_i - \sqrt{2}y_i) ]^2$ \\
 &  \\
$\gamma^2(f_i\rightarrow \eta\eta)$
&$[2\alpha^2z_i+2\sqrt{2}\beta^2y_i+r_2x_i]^2$ \\
 &  \\
$\gamma^2(f_i \rightarrow \pi\pi)$
&$3[z_i+r_2x_i]^2$ \\
 &  \\
$\gamma^2(f_i\rightarrow K\bar{K})$
&$4[\frac{1}{2}(z_i +\sqrt{2}y_i)+r_2x_i]^2$ \\
 & \\ \hline
\end{tabular}
\end{center}
\end{table}
\par
We then perform a $\chi^2$ fit based on the measured branching ratios.
The details of the fit are given in ref.~\cite{scalars}.
For this presentation the parameter $r_3$, used in
ref.~\cite{scalars},
has been set to zero.
As input we use the masses of the
$f_0(1500)$ and $f_0(1710)$.
In this way seven parameters, $M_G$,
$M_N$, $M_S$, $M_3$, $f$, $r_2$ and $\phi$
are determined from the fit.
The mass of the $f_0(1370)$ is not well established so we have left it as
a free parameter ($M_3$).
The
$\chi^2/NDF$ of the fit is 13.9/7 and the largest contribution comes
from the
$\frac{f_0(1710)\rightarrow \eta \eta}{f_0(1710)\rightarrow K \overline K}$
branching ratio which contributes 6.1 to the $\chi^2$.
\par
The mass parameters determined from the fit are
$M_G$~=~1446~$\pm$~16~MeV,
$M_S$~=~1664~$\pm$~9~MeV,
$M_N$~=~1374~$\pm$~28~MeV and
$M_3$~=~1248~$\pm$~31~MeV.
The output masses for $M_N$ and $M_S$
are consistent with the $K^*(1430)$ being in the nonet
and with the glueball
mass being at the lower end of the quenched lattice range.
The mass found for the $f_0(1370)$ (1256~$\pm$~31 MeV) is at the lower end
of the measured range for this state.
The pseudoscalar mixing angle is found to be
$\phi$~=~-25~$\pm$~4 degrees consistent with other determinations.
\par
The physical states $|f_0(1710)\rangle$, $|f_0(1500)\rangle$ and
$|f_0(1370)\rangle$ are found to be
\[
|f_0(1710)\rangle=0.42|G\rangle+0.89|S\rangle+0.17|N\rangle,
\]
\[
|f_0(1500)\rangle=-0.61|G\rangle+0.37|S\rangle-0.69|N\rangle,
\]
\[
|f_0(1370)\rangle=0.65|G\rangle-0.15|S\rangle-0.73|N\rangle.
\]

\par
Other authors have claimed that
$M_G>M_S>M_N$~\cite{Wein}.
This scenario is disfavoured as,
if in the fit we require
$M_G>M_S>M_N$, the $\chi^2$ increases to 57.
In any event,
we are cautious about such claims~\cite{Wein}
as they are likely to
be significantly distorted by the presence of a higher, nearby,
excited $n \bar{n}$ state ($N^*$) such that $M_{N^*}>M_G>M_S$:
the philosophy of
dominant mixing with the nearest neighbours would then lead
again to the ``singlet - octet - singlet"
scenario that we have found above.

\section{Predictions for production mechanisms}%4

Our preferred  solution has implications for the production
of these states
in $\gamma \gamma$ collisions, ${\rm p \bar{p}}$ annihilations,
in central ${\rm pp}$ collisions and in radiative $J/\psi$ decays.
These are interesting in that they
are consequences of the output and were not used as constraints.

\subsection{$\gamma \gamma$ production}%4.1

The most sensitive probe of flavours and phases is in $\gamma \gamma$
couplings. In the spirit of ref.~\cite{re:CFL}, ignoring mass-dependent
effects, the above imply
$
\Gamma(f_1(1710)\rightarrow \gamma\gamma):\Gamma(f_1(1500)\rightarrow
\gamma\gamma):\Gamma(f_1(1370)\rightarrow \gamma\gamma)=
3.8 \pm 0.9:6.8 \pm 0.8:16.6 \pm 0.9.
$
The $\gamma \gamma$
width of $f_0(1500)$ exceeding that of $f_0(1710)$ arises
because the glueball is nearer to the $N$ than
the $S$.
This shows how these $\gamma \gamma$
couplings have the potential to pin down the input pattern.

\subsection{${\rm p \bar{p}}$ production}%4.2

The production of the $f_0$ states in
${\rm p \overline p} \to \pi + f_0$ is expected
to be dominantly through the $n \overline n$ components of the
$f_0$ state, possibly through $gg$, but not prominently through
the $s \overline s$ components.
The above mixing pattern implies that
$\sigma({\rm p \overline p} \to \pi + f_0(1710)) <
\sigma({\rm p \overline p} \to \pi +  f_0(1370)) \sim
\sigma({\rm p \overline p} \to \pi +  f_0(1500))$
Experimentally~\cite{thoma} the relative production rates are,
${\rm p \overline p} \to \pi + f_0(1370) : \pi + f_0(1500)) \sim
1 : 1.$
and there is no evidence for the production of the $f_0(1710)$.
This would be natural if the production were via the
$n \overline n$ component.
The actual magnitudes would however be model dependent; at this stage we
merely note the consistency of the data with the results of the mixing
analysis above.
\subsection{Central production}%4.3

For central production,
the cross sections of well established
quarkonia in WA102 suggest that the
production of $s \overline s$ is strongly suppressed
relative to $n \overline n$.
The relative
cross sections for
the three states of interest here are
${\rm p p} \to {\rm pp} + ( f_0(1710):
 f_0(1500) :f_0(1370)) \sim 0.14:1.7:1. $
This would be natural if the production were via the
$n \overline n$ and $gg$ components.
\par
In addition,
the WA102 collaboration has studied the
production of these states as a function of
the azimuthal angle $\phi$, which is defined as the angle between the $p_T$
vectors of the two outgoing protons.
An important qualitative characteristic of these data is that
the $f_0(1710)$ and $f_0(1500)$ peak as
$\phi \to 0$ whereas the $f_0(1370)$ is more peaked
as $\phi \to 180$~\cite{WAphi}.
If the $gg$ and $n \overline n$
components are produced coherently as $\phi \to 0$ but out of phase
as $\phi \to 180$, then  this pattern of $\phi$ dependence and relative
production rates would follow; however, the relative coherence of
$gg$ and $n \overline n$ requires a dynamical explanation.
We do not have such an explanation and open this for debate.

\subsection{Radiative $J/\psi$ decays}%4.4
In $J/\psi$ radiative decays, the absolute rates depend
sensitively on the phases and relative strengths of the
$G$ relative to the $q \overline q$ component, as well as the
relative phase of $n\bar{n}$ and $s\bar{s}$ within the latter.
As discussed in
ref.~\cite{scalars}, based on the mixings found, we expect
that the
rate for $f_0(1370)$ will be smallest and that
the rate of $J/\psi \to \gamma f_0(1500)$ rate will be comparable to
$J/\psi \to \gamma f_0(1710)$.
\par
In ref.~\cite{dunwoodie}, the branching ratio of
BR$(J/\psi \rightarrow \gamma f_0)(f_0\rightarrow \pi\pi + K \bar{K})$
for the $f_0(1500)$ and $f_0(1710)$ is presented.
These can be used to show that~\cite{scalars}:
$J/\psi \rightarrow f_0(1500) : J/\psi \rightarrow f_0(1710)
= 1.0 : 1.1 \pm 0.4$
which is consistent with the prediction above based on
our mixed state solution.
\subsection{$\pi^-p$ and $K^-p$ production}%4.5
In these mixed state solutions,
both
the $f_0(1500)$ and $f_0(1710)$ have
$n \bar{n}$ and $s \bar{s}$ contributions and so
it would be expected that both would be produced in
$\pi^-p$ and $K^-p$ interactions.
The $f_0(1500)$
has clearly been observed in
$\pi^-p$ interactions and
there is also evidence for the production
of the $f_0(1500)$ in $K^-p \rightarrow K^0_SK^0_S \Lambda$.
\par
There is evidence for the $f_0(1710)$ in the reaction
$\pi^-p \rightarrow K^0_SK^0_Sn$,
originally called the $S^{*\prime}(1720)$.
One of the longstanding problems of the $f_0(1710)$ is that
in spite of its dominant $K \bar{K}$ decay mode it was
not observed in $K^-p$ experiments.
In ref.~\cite{lindenbaum} it was demonstrated that
if the $f_0(1710)$ had $J$~=~0, as it has now been found to have,
then the contribution in $\pi^-p$ and $K^-p$ are compatible.
One word of caution should be given here: the analysis
in ref.~\cite{lindenbaum} was performed with a $f_0(1400)$ not
a $f_0(1500)$ as we today know to be the case. As a further test of
our solution, it would
be nice to see the analysis of ref.~\cite{lindenbaum}
repeated with the mass and width of the
$f_0(1500)$ and the decay parameters of the $f_0(1710)$
determined by the WA102 experiment.
\section{Summary}%5
In summary,
based on the hypothesis that
the scalar glueball
mixes with the nearby $q \overline q$ nonet states,
we have determined the flavour content of the
$f_0(1370), f_0(1500)$ and $f_0(1710)$
by studying their decays into all pseudoscalar meson pairs.
The solution we have found
is also compatible with the relative production
strengths of the $f_0(1370), f_0(1500)$ and $f_0(1710)$
in ${\rm pp}$ central production, ${\rm p \bar{p}}$ annihilations
and $J/\psi$ radiative decays.

\end{document}